\documentclass[aps,prd,showpacs,showkeys,preprintnumbers,unsortedaddress,superscriptaddress,
twocolumn
]{revtex4-1}

\usepackage{amsmath,mathrsfs,amssymb}
\usepackage{graphicx}

\begin{document}

\title{A common origin of fermion mixing and geometrical CP violation, and its test through Higgs physics at the LHC}

\date{\today}

\author{Gautam Bhattacharyya}
\email{gautam.bhattacharyya@saha.ac.in}
\affiliation{Saha Institute of Nuclear Physics, 1/AF Bidhan Nagar, Kolkata 700064, India}
\affiliation{Fakult\"{a}t f\"{u}r Physik, Technische Universit\"{a}t
Dortmund D-44221 Dortmund, Germany}

\author{Ivo de Medeiros Varzielas}
\email{ivo.de@udo.edu}
\affiliation{Fakult\"{a}t f\"{u}r Physik, Technische Universit\"{a}t
Dortmund D-44221 Dortmund, Germany}

\author{Philipp Leser}
\email{philipp.leser@tu-dortmund.de}
\affiliation{Fakult\"{a}t f\"{u}r Physik, Technische Universit\"{a}t
Dortmund D-44221 Dortmund, Germany}

\keywords{CP violation; Flavor symmetries; Extensions of Higgs sector}

\pacs{11.30.Hv, 12.15.Ff, 12.15.Hh, 12.60.Fr}


\preprint{SINP/TNP/2012/12, DO-TH 12/28}

\begin{abstract}
We construct for the first time a flavor model, based on the smallest discrete symmetry $\Delta(27)$ that implements spontaneous CP violation with a complex phase of geometric origin, which can actually reproduce all quark masses and mixing data.
We show that its scalar sector has exotic properties that can be tested at the LHC.
\end{abstract}

\maketitle

In this Letter we present for the first time a viable flavor model of fermion masses and mixing that is able to account for all currently observed data, where we have employed the smallest group $\Delta(27)$ in which CP-violation of geometrical origin arises spontaneously during electroweak symmetry breaking through a unique complex vacuum expectation value (VEV). We require the Lagrangian to be invariant under the Standard Model (SM) gauge group, a discrete non-Abelian symmetry, and CP, which necessitates the introduction of additional scalars. We show that the extended scalar sector in this model contains novel phenomenology testable at the LHC.

The origin of CP violation is currently an open question in particle physics. In the SM, CP is broken due to complex Yukawa couplings and CP violation manifests itself in charged weak interactions through the Cabibbo-Kobayashi-Maskawa (CKM) matrix. Going beyond the SM it is possible to explore the origin of CP violation, and breaking CP spontaneously is particularly appealing\,\cite{Lee:1973iz,Branco:1979pv}.
In the framework of spontaneous CP violation, CP is a symmetry of the Lagrangian and therefore its parameters are real. CP violation can then arise from complex VEVs of the Higgs multiplets, provided the unitary transformation, $U$, given by
\begin{equation}
\langle H_i \rangle \longrightarrow \langle H_{i} \rangle^{\ast} = U_{ij} \langle H_{j} \rangle \,,
\label{eq:U}
\end{equation}
acting on the $H_i$ and relating the VEV to its complex conjugate, is \textit{not} a symmetry of the Lagrangian. If it is, then CP is conserved even though the VEVs are complex.

In this Letter, the calculable phase arising from geometrical CP violation (GCPV) is uniquely determined independently of the arbitrary parameters of the scalar potential. GCPV requires at least three Higgs doublets and a non-Abelian symmetry\,\cite{Branco:1983tn}. $\Delta(27)$ is known to be the smallest group for producing geometrical phases. In \cite{deMedeirosVarzielas:2011zw} this was generalized to larger groups obtaining the same calculable phases. Recently, several new phase solutions were advanced and expressed in terms of the number of scalars and the group \,\cite{Varzielas:2012pd}.

So far, viable models of fermion masses and mixing within the GCPV framework have not been constructed, although promising leading order structures have been proposed\,\cite{deMedeirosVarzielas:2011zw}. It has also been shown that the calculability of phases is robust and survives when the potential includes non-renormalizable terms\,\cite{Varzielas:2012nn}. Motivated by these previous works, we attempt here to produce for the first time the \textit{minimal} model of GCPV which can fit all data. For this purpose we base ourselves on a $\Delta(27)$ symmetry, a discrete subgroup of $\textsf{SU}(3)$ and the smallest group that leads to GCPV, and we add only the minimal amount of additional matter.

We assume, without any loss of generality, that the three Higgs doublets, $H_i$, transform as a $\Delta(27)$ triplet with an assignment of a $\mathbf{3}_{01}$ irreducible representation (irrep) with a \textit{lower} index. Their hermitian conjugates $H^{\dagger i}$ transform as the conjugate representation $\mathbf{3}_{02}$ with an \textit{upper} index, constituting the anti-triplet.
We now clarify our notation and illustrate some group properties. We denote the two relevant generators of the group as $c$ (cyclic permutation) and $d$ (diagonal phases). They operate as $c (H_1, H_2, H_3) \rightarrow (H_2, H_3, H_1)$, $c (H^{\dagger 1}, H^{\dagger 2}, H^{\dagger 3}) \rightarrow (H^{\dagger 2}, H^{\dagger 3}, H^{\dagger 1})$, and $d (H_1, H_2, H_3) \rightarrow (H_1, \omega H_2, \omega^2 H_3)$, $d (H^{\dagger 1}, H^{\dagger 2}, H^{\dagger 3}) \rightarrow (H^{\dagger 1}, \omega^2 H^{\dagger 2}, \omega H^{\dagger 3})$, where $\omega \equiv \mathrm{e}^{\mathrm{i} 2 \pi/3}$.
There are nine distinct singlet irreps $\mathbf{1}_{ij}$, where the subscript $\{ij\}$ denotes how they transform under the generators $c \mathbf{1}_{ij} = \omega^i \mathbf{1}_{ij}$, $d \mathbf{1}_{ij} = \omega^j \mathbf{1}_{ij}$.
More details about $\Delta(27)$ can be found in \cite{Luhn:2007uq,Ishimori:2010au}.

It was shown in \cite{Branco:1983tn} that the renormalizable scalar potential in the $\Delta(27)$ context can lead to a complex VEV of the type:
\begin{equation}
\langle H_i \rangle = v(\omega,1,1) \,,
\label{eq:VEV}
\end{equation}
that necessarily violates CP, as the corresponding $U$ (see Eq.~(\ref{eq:U})) is not a symmetry of the potential. We will revisit the scalar potential in greater detail later, but now it is important to focus on the Yukawa interactions.

We start with the quarks and recall the results of \cite{deMedeirosVarzielas:2011zw}.
In order to make invariant Yukawa terms some of the quarks must transform as triplet or anti-triplet under $\Delta(27)$\,\cite{Branco:1983tn}. We write the invariants symbolically as $Q H_i d^c$ and $Q H^{\dagger i} u^c$ (without explicit $\textsf{SU}(2)$ indices), with $Q$ the left-handed quark  doublets and $u^c$, $d^c$ as the up and down right-handed singlets. As described in \cite{deMedeirosVarzielas:2011zw}, by choosing $Q_i$ as a $\mathbf{3}_{01}$ we would necessarily require the $d^c_i$ to transform also as a $\mathbf{3}_{01}$. Instead, if $Q^i$ is a $\mathbf{3}_{02}$, the $u^{c i}$ is forced to be  a $\mathbf{3}_{02}$. The end result is at least one sector has a leading order Yukawa structure given by the $\Delta(27)$ invariant $\mathbf{3}_{0i}\otimes\mathbf{3}_{0i}\otimes\mathbf{3}_{0i}$. With the VEV in Eq.~(\ref{eq:VEV}), this structure leads to a mass matrix with three degenerate quark masses. We therefore conclude that $Q$ cannot be assigned as a triplet or an anti-triplet.
We are thus forced to choose instead $u^c$ and $d^c$ as $\Delta(27)$ triplets yielding $Q H_i d^{c j}$ and $Q H^{\dagger i} u^c_j$ with $Q$ as singlets. Both sectors have Yukawas arising as the $\Delta(27)$ invariants $\mathbf{1}_{ij} \otimes (\mathbf{3}_{01}\otimes\mathbf{3}_{02})$.
Although $\mathbf{3}_{01}\otimes\mathbf{3}_{02}$ results in 9 distinct singlets, the group properties are such that any $\mathbf{3}_{01}\otimes\mathbf{3}_{02} \rightarrow \mathbf{1}_{ij}$ with $i\neq0$ explicitly involves powers of $\omega$ (complex), so these possibilities are not allowed by CP invariance of the Lagrangian. To generate a renormalizable Yukawa interaction we are then restricted to assign $Q_1$, $Q_2$ and $Q_3$ each as one or the other of the three $\mathbf{1}_{0i}$ singlets. The remaining possibilities are then assigning all three $Q$ in the same singlet irrep or assigning two in the same, or all three $Q$ in different irreps. All three structures lead to mass matrices that have a special structure distinguished by rows. The choice of $Q$ as $\mathbf{1}_{00}$, $\mathbf{1}_{01}$ or $\mathbf{1}_{02}$ forces the respective $H_i d^{c j}$ or $H^{\dagger i} u^{c}_j$ product to be $\mathbf{1}_{00}$, $\mathbf{1}_{02}$ or $\mathbf{1}_{01}$ respectively, which essentially amounts to a shift in the position of the $\omega$ in the mass matrix. More explicitly the corresponding down mass matrix looks like:

\begin{equation}
\tilde{M}_d = v \begin{pmatrix}
	y_{1} \omega & y_{1} & y_{1} \\
	y_{2} & y_{2} \omega & y_{2} \\
	y_{3} & y_{3} & y_{3} \omega 
\end{pmatrix}
\end{equation}
and the associated up quark mass matrix looks very similar ($\omega^2$ instead of $\omega$ and the second and third rows swapped).
Conversely, if $Q_1$, $Q_2$, and $Q_3$ are assigned to $\mathbf{1}_{00}$, $\mathbf{1}_{00}$, and $\mathbf{1}_{02}$ respectively, we get:
\begin{equation}
M_d = v \begin{pmatrix}
	y_{1} \omega & y_{1} & y_{1} \\
	y_{2} \omega & y_{2} & y_{2} \\
	y_{3} & y_{3} & y_{3} \omega 
\end{pmatrix}
\end{equation}
We recall that due to the explicit CP invariance of the Lagrangian, the Yukawa couplings are all real, and the phase appears only through the complex VEV.
At this point it is instructive to show the hermitian matrices $M M^\dagger$:

\begin{equation}
\tilde{M}_d \tilde{M}_d^{\dagger}= 3 v^2 \begin{pmatrix}
	y_{1}^2 & 0 & 0 \\
	0 & y_{2}^2 & 0 \\
	0 & 0 & y_{3}^2
\end{pmatrix}
\end{equation}
Vanishing off-diagonal entries follow from $1+\omega+\omega^2=0$.

Finally, for $M_d$ we have: 
\begin{equation}
M_d M_d^{\dagger}= 3 v^2 \begin{pmatrix}
	y_{1}^2 & y_{1} y_{2} & 0 \\
	y_{1} y_{2} & y_{2}^2 & 0 \\
	0 & 0 & y_{3}^2
\end{pmatrix}
\end{equation}
Note that the determinant of this structure is zero but it has two non-vanishing masses.
The last choice, all generations of $Q$ in the same singlet irrep leads to a rank 1 structure with a single non-vanishing mass.
Another relevant observation is that the complex phase is entirely absent in all these hermitian structures.

In order to obtain a viable CKM matrix it is necessary to generate additional off-diagonal terms. The minimal way to do this is to add a gauge singlet scalar that is a non-trivial $\Delta(27)$ singlet, which we denote as $\phi$. Without any loss of generality we place $\phi$ in the irrep $\mathbf{1}_{01}$. This enables a new non-renormalizable Yukawa coefficient per row, associated with terms of the type $Q H_i d^{c j} \phi$. For $Q_1$, $Q_2$, and $Q_3$ in $\mathbf{1}_{00}$, $\mathbf{1}_{00}$, and $\mathbf{1}_{02}$ respectively, we have to add to $M_d$ the corresponding mass matrix:
\begin{equation}
M_{\phi} = v \begin{pmatrix}
	y_{\phi1} & y_{\phi1} \omega & y_{\phi1} \\
	y_{\phi2} & y_{\phi2} \omega & y_{\phi2} \\
	y_{\phi3} \omega & y_{\phi3} & y_{\phi3} 
\end{pmatrix}
\label{eq:yphi}
\end{equation}
From the interference $M_{d} M_{\phi}^\dagger + M_{\phi} M_{d}^\dagger$ we obtain the required off-diagonal entries whereas the effect of $M_{\phi} M_{\phi}^\dagger$ can be absorbed within the structure of $M_d M_{d}^{\dagger}$. 

A complex phase in the CKM matrix requires that the hermitian matrices of the $M M^\dagger$ type are complex, which is not the case up to now. To preserve the complex phase in the hermitian matrices requires a further augmentation. The minimal possibility is to consider the non-renormalizable interactions that contain higher powers of $H$ e.g.~$Q H_i d^{c j} (H_k H^{\dagger l})$. The only non-trivial structure that we extract from the last non-renormalizable combination is:
\begin{equation}
M_{H} = v \begin{pmatrix}
	y_{H1}  & y_{H1} \omega^2 & y_{H1} \omega^2 \\
	y_{H2} & y_{H2} \omega^2 & y_{H2} \omega^2 \\
	y_{H3} \omega^2 & y_{H3} \omega^2 & y_{H3}
\end{pmatrix}
\end{equation}
where the identity $1+\omega+\omega^2=0$ was used and the existing coefficients were redefined to absorb similar entries in the mass matrix. From the interference $M_{d} M_{H}^\dagger + M_{H} M_{d}^\dagger$ we obtain the phases that enable complex CKM elements, whereas both $M_{\phi} M_{H}^\dagger + M_{H} M_{\phi}^\dagger$ and $M_{H} M_H^\dagger$ give structures that do not qualitatively change the analysis. The essential point is that the presence of $M_H$ is crucial to generate the phase.

\begin{figure}
\begin{center}
 \includegraphics{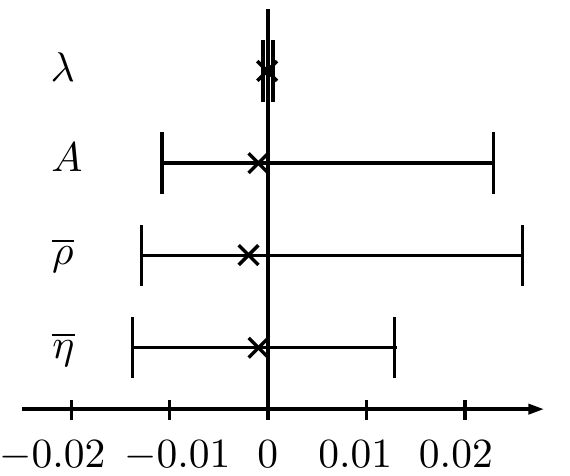}
 \caption{The experimental spread of the Wolfenstein parameters $\lambda$, $A$, $\bar{\eta}$ and $\bar{\rho}$ around their central values \cite{Beringer:1900zz}. Crosses denote our model values. \label{fig:Wolfenstein}}
\end{center}
\end{figure}

Note that $M_{\phi}$ and $M_{H}$ are the minimal mandatory additions that are necessary for a perfect fit to the existing data.
Following the above chain of arguments, we finally write the relevant Lagrangian, explicitly showing  the $\Delta(27)$ multiplet indices, as:
\begin{equation}
\mathcal{L} = Q \left(H^{\dagger i} u^{c}_j + H_i d^{cj} + H_i d^{cj} \phi + H_i d^{cj} (H_k H^{\dagger l})  \right) \,.
\end{equation}
In fact we found that the only choice that favorably accounts for the precision flavor data is when $Q_1$, $Q_2$ and $Q_3$ are chosen as $\mathbf{1}_{00}$, $\mathbf{1}_{00}$ and $\mathbf{1}_{02}$ respectively. Concerning the up quark sector, $M_u M_u^\dagger$ can be considered to be diagonal, and we need only one additional non-renormalizable Yukawa in order to generate the small up quark mass. In Fig.~\ref{fig:Wolfenstein} we show that with this choice we can successfully reproduce the Wolfenstein parameters from \cite{Beringer:1900zz} (the values we obtained are presented in the right column below):
\begin{equation}
\label{eq:wolfenstein}
\begin{aligned}
 \lambda^\text{exp} &= 0.22535 \pm 0.00065 & \lambda &= 0.22534, \\
 A^\text{exp} &= 0.811 \begin{matrix}
 +0.022\\
 -0.012
 \end{matrix} & A &= 0.810,\\
 \bar{\rho}^\text{\;exp} &= 0.131 \begin{matrix}
  +0.026\\
  -0.013
 \end{matrix} & \bar{\rho}  &= 0.129,\\
 \bar{\eta}^\text{\;exp} &= 0.345 \begin{matrix}
 +0.013\\
 -0.014
 \end{matrix} & \bar{\eta} &= 0.344.
\end{aligned}
\end{equation}

What about the leptons? This sector is experimentally less constrained than the quark sector. The possible invariants depend on what is responsible for the generation of neutrino masses e.g. the type of seesaw mechanism, as discussed in \cite{deMedeirosVarzielas:2011tp}. In addition to the structures that fit the quark sector, other representation choices can also work in the lepton sector. A leptonic model based on the $\mathbf{3}_{0i}\otimes\mathbf{3}_{0i}\otimes\mathbf{3}_{0i}$ invariants in $\Delta(27)$ has been discussed in \cite{Ma:2006ip} (for the $A_4$ group see the detailed analysis in \cite{Toorop:2010ex,*Toorop:2010kt}).

We now turn our attention to the scalar potential which contains the $\Delta(27)$ triplet $H_i$ as well as $\phi$.
The full renormalizable potential is (recalling that all couplings are real):
\begin{multline}
 V(H,\phi) = m_1^2 \left[ H_1 H_1^\dagger \right]+
m_2^2 \phi \phi^\dagger + 
m_3 (\phi^3 + \text{h.c.})\\
+\lambda_1 \left[(H_1 H_1^\dagger)^2\right]+
\lambda_2 \left[H_1 H_1^\dagger H_2 H_2^\dagger \right]+
\lambda_3 \left[H_1 H_2^\dagger H_1 H_3^\dagger
+\text{h.c.}\right]\\
+\lambda_4 (\phi\phi^\dagger)^2+
\lambda_5 \left[\phi (H_1 H_2^\dagger)+\text{h.c.}\right]+
\lambda_6 \left[\phi \phi (H_1 H_3^\dagger)+\text{h.c.}\right]\,,
\end{multline}
where the square brackets represent also the cyclic permutations on the $\Delta(27)$ indices which we do not explicitly show.
The geometrical phase solution in Eq.~(\ref{eq:VEV}) is not affected by $\phi\phi^\dagger$. When $\lambda_5$ and $\lambda_6$ are small, Eq.~(\ref{eq:VEV}) holds, and otherwise one can add a $Z_4$ symmetry acting on $\phi$ to trivially enforce them to vanish (in this case Eq.(\ref{eq:yphi}) arises from a $\phi^4$ insertion instead of $\phi$, all conclusions remaining unchanged).
For illustration we display only the CP-even scalar components (in this class of models, one can separately identify scalars and pseudoscalars \cite{Lavoura:1994fv}). Following the minimization of the potential and determination of the mass eigenvalues, we observed these features: $(i)$ The $\phi$ field is much heavier (beyond $1$\,TeV) and decouples from the $\textsf{SU}(2)$ doublets. More specifically, the mass of $\phi$ is determined by $\lambda_{\{4,5,6\}}$, while those of $h_{\{a,b,c\}}$ are controlled by $\lambda_{\{1,2,3\}}$.
$(ii)$ The physical scalars $h_a, h_b$ and $h_c$ mix in a very specific way as witnessed in \cite{Ma:2010gs,Bhattacharyya:2010hp,Cao:2011df,Bhattacharyya:2012ze} primarily in the $S_3$ context. The scalar mass squared matrix has the structure
\begin{equation}
	\begin{pmatrix}
		A & B & B\\
		B & C & D\\
		B & D & C
	\end{pmatrix},
\end{equation}
which leads to one physical scalar $h_a$ that is orthogonal to the other scalars and has no $h_a VV$-type gauge couplings ($V=W,Z$). Its Yukawa couplings to up- and down-type quarks are strongly suppressed, except that the $h_a ct$ and $h_a uc$ couplings are about $0.45$. The other physical scalars, $h_b$ and $h_c$, have almost SM-like gauge and Yukawa couplings.
 
\begin{figure}
\begin{center}
 \includegraphics{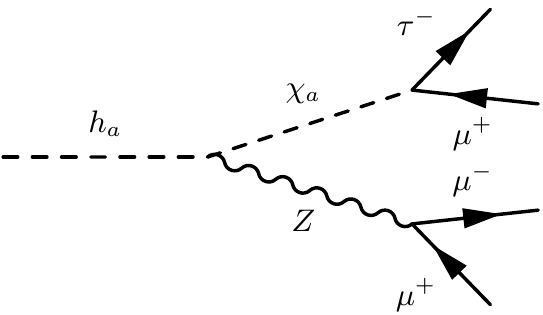}
 \caption{Example of a decay mode of the exotic scalar $h_a$ that can be tested at the LHC. \label{fig:Feynman}}
\end{center}
\end{figure}

Adjusting the scalar potential couplings, two viable scenarios can be identified: (I) There is only one light scalar, $h_b$, that plays the role of the SM-like Higgs found near $125$\,GeV. In this case all other scalars are beyond the current exclusion range of the LHC. This may be considered as a decoupling limit which reproduces almost SM-like scalar structure.  
(II) A scenario which has richer collider consequences emerges when the exotic scalar $h_a$ is light enough to be produced (either through $h_a uc$, or through top or heavy scalar decays) at the LHC. Under the reasonable assumption that $m_\phi$ is greater than $1$\,TeV or so we can obtain the following analytic relations:
  \begin{align}
  m_{h_a}^2 &= \frac{2}{3} \left(2 \lambda_1 v^2-2 \lambda_2 v^2+3 \lambda_3 v^2\right),\\
  m_{h_{c/b}}^2 &= \frac{1}{6} \bigl(5 \lambda_1 v^2+4 \lambda_2 v^2
  \pm\sqrt{3} \bigl[v^4 \bigl(3 \lambda_1^2+8 \lambda_1
   \lambda_2\nonumber\\&\quad-16 \lambda_1 \lambda_3+16 \lambda_2^2-64 \lambda_2
   \lambda_3+64 \lambda_3^2\bigr)\bigr]^\frac{1}{2} \bigr).
  \end{align}
It is possible to adjust the potential couplings $\lambda_i$ to yield $m_{h_a}$ around or perhaps slightly larger than the mass $125$\,GeV of the SM-like $h_b$, with $h_c$ heavier than $600$\,GeV.
In this case, a spectacular decay channel opens through $h_a\to\chi_a Z$, fixing $m_{\chi_a} \sim 20$\,GeV, with subsequent decays of the pseudo-scalar $\chi_a $ to charged leptons of different flavors (e.g. $\mu \tau$) and of the $Z$ boson to leptons -- see Fig.~(\ref{fig:Feynman}). There is enough freedom in the lepton sector to boost this $\chi_a$ coupling, which may generate a sizable branching ratio in this channel; however, a more specific prediction requires a detailed numerical study of the lepton Yukawa sector which we do not delve into here.

In both scenarios (I) and (II) there are quite a few heavy scalars above the current LHC limit of  650 GeV or so, but their relative heaviness compared to the SM Higgs is not a result of fine-tuning of parameters as for each heavy state there exists a reasonably independent combination of $\lambda_i$-type couplings which simply has to be set to a higher value. We have verified this numerically.
If the LHC bound goes up, we have to accordingly increase the maximum allowed value of some $\lambda_i$ beyond $\pi$, and thus depending on how far the experimental limit is pushed up, we would have to consider higher values of $\lambda_i$ e.g. up to maximum allowed value of $2 \pi$ to go over $1$\,TeV.
Note that a proliferation of scalar states below $1$\,TeV in both scenarios (I) and (II), all coupling to SM gauge bosons, would affect the energy dependence of longitudinal gauge boson scattering. This energy dependence might be different from the SM expectation due to the presence of the extra scalars, whose quantitative impact may be probed at the high luminosity option of the LHC depending on their masses and couplings. A detailed analysis is beyond the scope of this Letter.

In summary, we have for the first time reproduced the CKM mixing matrix in a minimal $\Delta(27)$ flavor model, which is the smallest group where one can implement spontaneous CP violation of geometrical origin. Since quark mixing can be tested in several different independent channels, to reproduce the CKM matrix in a minimal scenario is often more difficult than fitting the lepton mixing. Within the framework of a large class of discrete symmetries it is usually difficult to exclude different choices of representations from data. But our scenario is quite falsifiable, in the sense that only two choices broadly worked, out of which only one set of matter and Higgs representations fits the ever growing precision of flavor data. The scalar sector of the model inherits enough symmetries of the flavor group which induce exotic scalar decays into multi-lepton of different flavors, constituting a smoking gun signal of the model testable at the LHC.

\begin{acknowledgments}
We thank Lu\'is Lavoura for his comments on the manuscript.
G.B. acknowledges DFG support through a Mercator visiting professorship, grant number INST 212/289-1, and hospitality at TU Dortmund.
The work of I.d.M.V. was supported by DFG grant PA 803/6-1 and partially through PTDC/FIS/098188/2008.
The work of P.L. was supported by the Studienstiftung des deutschen Volkes.
\end{acknowledgments}
\bibliography{refs}

\end{document}